# A Semi-supervised Sensing Rate Learning based CMAB Scheme to Combat COVID-19 by Trustful Data Collection in the Crowd

Jianheng Tang[a], Kejia Fan[a], Wenxuan Xie[a], Luomin Zeng[b], Feijiang Han[a], Guosheng Huang[c,*], Tian Wang[d], Anfeng Liu[a,*], Shaobo Zhang[e]

[a] School of Computer Science and Engineering, Central South University, Changsha, 410083, China

[b] School of Civil Engineering, Central South University, Changsha, 410083, China

[c] School of computer Science and Engineering, Hunan First Normal University, Changsha, 410205, China

[d] Department of Artifficial Intelligence and Future Networks, Beijing Normal University & UIC, Zhuhai, Guangdong, China

[e] School of Computer Science and Engineering of the Hunan University of Science and Technology, Xiangtan, 411201, China

---

| ARTICLE INFO | ABSTRACT |
|---|---|
|  | The recruitment of trustworthy and high-quality workers is an important research issue for MCS. Previous studies either assume that the qualities of workers are known in advance, or assume that the platform knows the qualities of workers once it receives their collected data. In reality, to reduce costs and thus maximize revenue, many strategic workers do not perform their sensing tasks honestly and report fake data to the platform, which is called False data attacks. And it is very hard for the platform to evaluate the authenticity of the received data. In this paper, an incentive mechanism named Semi-supervision based Combinatorial Multi-Armed Bandit reverse Auction (SCMABA) is proposed to solve the recruitment problem of multiple unknown and strategic workers in MCS. First, we model the worker recruitment as a multi-armed bandit reverse auction problem and design an UCB-based algorithm to separate the exploration and exploitation, regarding the Sensing Rates (SRs) of recruited workers as the gain of the bandit. Next, a Semi-supervised Sensing Rate Learning (SSRL) approach is proposed to quickly and accurately obtain the workers' SRs, which consists of two phases, supervision and self-supervision. Last, SCMABA is designed organically combining the SRs acquisition mechanism with multi-armed bandit reverse auction, where supervised SR learning is used in the exploration, and the self-supervised one is used in the exploitation. We theoretically prove that our SCMABA achieves truthfulness and individual rationality and exhibits outstanding performances of the SCMABA mechanism through in-depth simulations of real-world data traces. |

## 1. Introduction

Mobile CrowdSensing (MCS) has emerged as a promising data sensing and collection paradigm, monitoring the real world through the vast amount of data obtained from embedded sensors in mobile devices [1]–[3]. The MCS typically consists of three organic components [4]–[6]: task requesters, data platform and workers. The requesters order their tasks to the platform [7], [8], and the platform publishes these tasks and recruits workers to complete them [9]–[11]. The workers are people who carry smart devices embedded with rich sensors, such as smartphones, smartwatches or others [12]–[14]. And they can report data to the platform for rewards through participatory sensing [15]–[17]. According to a recent survey from International Data Corporation (IDC) [18], 41.6 billion IoT devices will be connected in 2025, producing 79.4 Zettabytes (ZB) of data. Thanks to such a large number of sensing devices [19], [20], a variety of MCS-based applications have emerged in recent years, including VTrack [21], NoiseTube [22], WeatherLah [23], and many more [24], [25]. For example, with the spreading of COVID-19, more and more people are suffering from physical pains and mental pressures [26], [27]. MCS can act as an effective technology for detecting COVID-19 and blocking its spreading [27], [28]. To this end, AI-enabled data sensing and data collection technologies could be used to monitor the physical status of the crowd, based on the received multimodal data such as crowd flow trajectory, images and sounds [29]. We can detect and predict the extent and trend of COVID-19 transmission in time by analyzing the massive quantity of data obtained, enabling us to take appropriate measures to prevent or interrupt the spread of COVID-19, which can significantly decrease the risks of human infection [30].

The success of MCS applications depends on the massive data contributed by the workers [31]–[33]. In completing tasks, workers will have some costs, such as transportation fees, time-consuming, communication overhead, and device loss. To encourage more workers to participate in the MCS, requesters need to give some rewards to the contributing workers to compensate for their costs [34]-[36]. Workers expect to be rewarded significantly for giving sensing data, yet the platform typically wants high-quality data at a minimum cost due to budgetary constraints [37]-[38]. The auction is a game mechanism often used as an effective tool to solve the above problem in the MCS system, for the smooth operation of the applications residing on the platform [26], [28]. In auction-based MCS, the platform seeks to recruit trustworthy workers with high qualities to carry out tasks [4] to ensure the performance of the applications [9], which is a crucial issue of the research [36].

---

＊Corresponding author.

*E-mail address: tangent-heng@csu.edu.cn (J. Tang), kejiafan@csu.edu.cn (K Fan), 8208200422@csu.edu.cn (W. Xie), luominzeng@csu.edu.cn (L. Zeng), feijianghan@gmail.com (F. Han), huanggs@hnfnu.edu.cn (G. Huang), tianwang@bnu.edu.cn (T. Wang), afengliu@mail.csu.edu.cn (A. Liu), shaobozhang@hnust.edu.cn (S. Zhang).*

However, recruiting trustworthy workers with high qualities from a large number of workers is quite challenging [39]-[41]. While many studies have been conducted on recruiting high-quality workers to perform tasks in MCS [42], [43], there is still much room for further researches in this area.

The first type of research assumes that the platform knows in advance the qualities of the workers [10], [44]. Based on this assumption, the selection of workers becomes an optimal worker selection problem under the platform budget constraint [37], [38]. However, the actual situation in MCS is different from this. There are a large number of workers in the MCS, but the platform does not know the sensing abilities and qualities of these workers in advance, so recruiting workers remains a major problem. Previous researchers have named this kind of problem the Unknown Worker Recruitment (UWR) problem [4], [12]. The second type of research can determine the qualities of workers based on data submitted by workers. To solve the UWR problem, Multi-Armed Bandit (MAB) based worker recruitment is used as an effective mechanism [4], [34], [37]. The current MAB approaches mainly use the strategy of exploration and exploitation [12], [34], [37]. In MCS, the number of workers is enormous. However, there are significant differences in their credibility and sensing qualities, and also, the platform is unaware of these workers' attributes. Therefore, it is necessary for exploration to select some workers and ask them to submit data in order to obtain and evaluate the qualities of their submitted data and their abilities to complete the tasks. After obtaining workers' sensing abilities, the platform is able to exclude low-quality workers from the candidate workers and select high-quality workers for data collection, thereby boosting the system's quality [12], [34], [37]. In this way, exploration is the process of selecting workers to test their abilities without knowing any information in advance. At the same time, the platform needs to pay for recruiting workers, and if the qualities of the recruited workers are low, the quality of the application will be affected. Exploitation is the utilization of the platform's information about workers' sensing qualities to select high-quality workers. MAB is a combination of exploration and exploitation to obtain better results [12], [34], [37].

The application of the MAB mechanism in MCS has been extensively studied [12], [34], [37]. However, according to the investigation, all of the proposed MAB mechanisms are based on the assumption that the platform can determine the sensing qualities of workers after obtaining the data [12], [34], [37]. In fact, the MCS platform usually cannot directly judge the sensing qualities of workers, or even whether the reported data are truthfully perceived or maliciously falsified. Therefore, this assumption is incorrect in MCS, making the proposed MAB mechanism difficult to apply in practice. In MCS, verifying the authenticity of worker-reported data is a challenging issue known as Information Elicitation Without Verification (IEWV) problem [45]. And it is common in many practical MCS applications [46]-[48]. Since the authenticity of the reported data is a prerequisite for the correct operation of applications, the current MAB mechanism deserves further study.

Therefore, to improve the data quality, it is necessary to obtain the Ground Truth Data (GTD) first and solve the IEWV problem [45]. In solving this problem, researchers have proposed some truth data discovery methods to ensure that the data received by the MCS platform are truthful [49]-[51]. Most of these methods are based on the assumption that most workers are trustworthy and the data given by them is truthful. Based on this assumption, redundant sampling methods are widely used in the proposed truth data discovery methods. These methods engage redundant workers for data collecting, then use the Mean method [40], Median method [40], and Weighted Mean method [47] to infer the truth. This method can obtain data close to GTD under certain conditions. However, the platforms using these methods are also vulnerable to attack. For instance, in the Mean method [40], if individual workers report data that deviate from the true data, the data obtained by the platform will diverge significantly from the GTD. Moreover, if multiple malicious workers make a coordinated attack, it is easy for the platform to get the attackers wanted results. Therefore, considering the presence of malicious or untrustworthy workers [52] and selecting trustworthy workers [53] has become a critical issue in MCS systems [46]-[48]. In the case of COVID-19, incorrect data might cause significant harm, either by spreading the virus [28], putting people's health at risk, or wasting more health resources [29], [30]. In this case, it is urgent to ensure that the platform selects trustworthy workers for truthful data [13], [39], [40].

Currently, the IEWV problem is being addressed primarily to validate the true data obtained [45], and these studies deserve exploratory research. The proposed MAB mechanism assumes that the platform knows the sensing qualities of the workers after it obtains the workers' data, which is an invalid assumption [12], [34], [37]. Thus, it is urgent to integrate these two problems well in order to recruit trustworthy workers to obtain truth data for MCS applications without the above invalid assumptions. Such a problem is a huge challenge that has not been properly addressed in the literature until now. To tackle such a problem, we propose an incentive mechanism based on Semi-supervision based Combinatorial Multi-Armed Bandit reverse Auction (SCMABA) scheme to recruit high-quality and low-bid workers in MCS. The SCMABA mechanism consists of two main parts. The first part is an effective workers' Sensing Rates (SRs) learning approach to obtain workers' sensing abilities. The SRs are used to characterize workers' probabilities of truthfully completing tasks according to quality requirements. If a worker always reports the data truthfully, his SR will be high; if a worker tends to report fake data, his SR will be low. The second part is an effective worker selection mechanism based on SRs combining the Combinatorial Multi-Armed Bandit (CMBA) and the reverse auction mechanism. It effectively addresses the shortcomings of the previous assumptions of the MAB mechanism. The importance and originality of our work lie in that it reveals some interesting and even counterintuitive findings mainly by discarding the two strong assumptions in previous studies. The first one is that we discard the assumption in previous studies that the platform knows the data quality once it obtains it. The second one is that we use Ground Truth Data (GTD) as the data benchmark to obtain the SRs, abandoning the assumption that most of the submitted data are truthful in previous studies. Our research provides fresh insights into understanding the complexity of MCS. These findings can boost forward-looking and strategic planning solutions. To summarize, the main innovations of this work are as follows:

(1) First, we abandon the unreasonable assumption that the platform can determine the sensing quality of workers after obtaining data from previous studies, and a Semi-supervised Sensing Rate Learning (SSRL) approach is proposed to quickly and accurately obtain the workers' SRs. Specifically, the SSRL approach is divided into 2 phases: supervision and self-supervision. By judging the authenticity of workers' submitted data, we can quickly and accurately calculate the SR of each worker.

(2) We designed a new Semi-supervision based Combinatorial Multi-Armed Bandit reverse Auction (SCMABA), which organically combines the SR acquisition mechanism with multi-armed bandit reverse auction. Specifically, in SCMABA, we use workers' SR as the gain of the multi-armed bandit, thus dividing the recruitment process of workers into two phases: exploration and exploitation. In the exploration and exploitation phases, the platform uses the supervised and self-supervised SR learning methods to calculate the SRs of workers, respectively.

(3) We design a UCB-based greedy algorithm to recruit workers and compute payments for SCMABA. In addition, we theoretically prove that our SCMABA mechanism can achieve truthfulness, individual rationality and computational efficiency in each recruitment round and then analyze the regret of the mechanism.

(4) We demonstrate significant performances of the SCMABA mechanism through extensive simulations of real-world data. The results show that SCMABA increases the platform revenue by at least 9.34% and decreases the platform regret by at least 18.59%.

The rest of this paper is organized as follows. Section 2 introduces related works. The system model, problem and definitions are presented in Section 3. In Section 4, we propose the SCMABA mechanism. Then, Section 5 provides the theoretical and experimental analysis. Finally, the conclusion and future work are given in Section 6.

## 2. Related work

With the development of microprocessor technology, an increasing number of sensing devices (e.g., smartphones and smartwatches) have emerged, and their number has exceeded 20 billion [54], [55], [56]. They are used in various application scenarios because of their powerful abilities in large-scale and long-time data sensing [57], [58]. In particular, the development of contact-free sensing technology has made data sensing and acquisition more convenient [59]. With the emergence of portable and highly integrated commodity-grade WIFI, radar, and sonar-style devices [60], numerous applications such as human health monitoring and environmental monitoring can be realized [59]. For example, COVID-19 is highly infectious [27], [29], and if non-contact sensing is used in combination with MCS, large-scale, low-cost, and secure data acquisition can be achieved [61], [62]. Due to its great application prospects, MCS research has attracted much attention [51], [52]. In MCS, there are three important components [1]-[3]. (1) Task requesters. Task requesters submit task requests to the platform, and also give the payment for completing the task [3]-[6]. (2) MCS Platform. The MCS platform is the most important component of MCS. For one thing, it accepts task requests from task requesters [7]-[9]. For another, tasks are posted on the platform, and the platform recruits low-priced, high-quality workers to complete them through auctions or other means. Researchers believe that high-quality data and the low bid of workers can bring big revenue to platforms and applications [12], [34], [37]. (3) Workers. The workers generally refer to people who hold sensing mobile devices with embedded sensors [34], [35], [61]. Workers are informed of tasks posted by the platform, then bid for the tasks they are interested in. After winning the bids, they are supposed to complete the tasks and submit the data truthfully. Finally, they will the rewards to cover the cost of resources, effort, and time consumed by completing the tasks themselves [31]-[33]. There are two modes of workers' participation in the task. One is participatory sensing, which requires workers to actively change their behavior, go to the task location at the specified time, and complete the sensing task, which is costly for workers [34], [35], [61]. Another common sensing model is opportunistic sensing, which does not require workers to change their behavior, and when they pass by a task location, the corresponding sensing task is completed incidentally [6], [8], [10]. In this approach, workers do not need to move specifically to complete the task but merely to piggyback on it, which makes it less costly. Moreover, the sensing tasks can be divided into real-time and non-real-time tasks depending on their completion time requirements. While real-time tasks are sensitive to task completion time, non-real-time tasks usually have a wide range of task completion times. For example, non-real-time MCS is often used for long-term continuous sensing tasks, such as gathering a group of volunteers with sensing devices to perform fine-grained continuous (e.g., months) sensing of air quality in a city [13]-[15].

This paper focuses on obtaining the qualities of workers and recruiting high-quality workers to complete the tasks [14], [62]. There are two key aspects involved as follows: (1) How to evaluate and obtain the qualities of workers' sensing data, i.e., the abilities of workers to sense data [14], [62]. Only by knowing the workers' sensing abilities (or quality) can the platform effectively recruit high-quality workers from the candidate workers and thus synthesize high-quality applications [21]-[23]. Data contributed by workers are the basis of MCS applications and its data quality determines the quality of applications [14], [62]. (2) How to trade-off exploration and exploitation in MCS so that the platform can both explore to obtain the workers' sensing abilities and also utilize the obtained information to select high-quality workers at a minimum cost. This will help the platform get as much high-quality data as possible with a limited budget to fulfill the mission of MCS.

The first key aspect we address is how to evaluate and obtain the sensing abilities of workers. Currently, there are several studies on the evaluation and acquisition of workers' sensing qualities. These studies are divided into the following categories of approaches, depending on the starting point of the evaluation:

(1) One type of study is to assume that the platform already knows the data qualities of all workers before selecting workers, allowing the platform to select high-quality workers and get high-quality data using various optimization strategies [9]-[11]. Knowing the sensing qualities of all workers in advance is only possible with any method, and such approaches are tough to apply in practice. However, this assumption can significantly simplify the problem of selecting workers for the platform. Based on this assumption, the platform can select high-quality workers to complete the task, but it is often not so easy in practice. A few studies in this area are given below: In one case, workers are assumed to arrive randomly in a real-time task. This case also challenges the platform on worker selection because workers' trajectories are random. It is probable that workers who arrive first have low sensing qualities, and those who arrive later have higher sensing qualities. The platform needs to predict the qualities of the workers arriving later. Thus, as each worker arrives, the platform must decide whether to pick up or drop off the worker [63]. If the worker is selected, the workers arriving later may have higher sensing qualities. In contrast, if the worker is dropped, it is possible that the workers arriving later will have lower sensing qualities or no workers will arrive in a limited time. Moreover, the worker selection problem becomes more complex if the sensing qualities of the workers are added to the workers' bids [63]. Because in this case the optimization objectives for worker selection change from 1-dimension to 2-dimension, and the complexity increases dramatically. For example, earlier workers may have higher sensing qualities and higher bids, while some have lower sensing qualities and lower bids [63]. Because it is still being determined whether there may be workers with high sensing qualities but low bids later, making the strategy design more difficult. In addition, time-sensitive tasks further complicate the problem because they generate values that vary with time [63]. For example, for urgent abrupt events such as a communication jam, the earlier the data is reported, the higher the value will be obtained. Early workers are more likely to be selected in this situation, even if their sensing qualities are low. Because as time goes by, the value generated by completing the task is discounted. Thus, the problem becomes more complex when considering the time, sensing ability, and bid together [63]. In fact, even when the sensing qualities of workers are known, it is still challenging to design an exemplary algorithm for the platform selection of workers, especially for MCS. For instance, in environmental monitoring applications, the network is usually divided into numerous grids, each grid representing a small task and only a limited number of workers are required to be recruited for cost reasons [10], [11]. In MCS, recruiting more workers does not significantly improve the quality of the application, but rather increases the cost of the platform. However, the number of workers willing to perform data collection varies for geographically different grids, which causes difficulties in recruiting workers [10], [11]. For example, Ren et al. [10] argue that in a smart city, the number of workers willing to complete the sensing tasks are particularly high in the central areas of the city, but relatively low in the peripheral areas of the city. So effective incentives should be used to motivate workers in peripheral areas to participate in sensing. In their study, they transformed such a problem into an optimization problem of pricing different regional tasks to improve the quality of the application with a certain budget. Their approach is to give lower rewards to areas where abundant workers are willing to participate and higher rewards to tasks in areas where workers are scarce, so that the data collected for each grid meet the requirements while minimizing the cost [10], [11].

(2) Another type of study abandons the assumption that the qualities of workers are known a priori. Instead, it assumes that the qualities of workers' submissions are available if the worker submits the data [12], [34], [37]. In this case, the selection of workers requires the following method: For one thing, it constantly explores the qualities of its workers' data, and for another, it prioritizes the selection of high-quality workers after obtaining information about them. In this case, if the number of workers is vast and frequently appears and disappears in the dynamic network, the platform can only obtain the quality status of some workers because of the cost and time. Thus, the platform should explore the qualities of unknown workers while selecting workers with high qualities. The results obtained by the platform could be more optimal but in line with the general situation of the objective world [12], [34], [37]. The above problem is the Unknown Worker Recruitment (UWR) problem [13], [39], [40]. The Combinatorial Multi-Armed Bandit (CMAB) scheme has proven to be an effective method for solving this type of problem [12], [34], [37]. The main function of the MAB mechanism is reflected in the design of the Upper Confidence Bound (UCB) algorithm, whose essence is how to select workers. The principle of the UCB algorithm is that if the sensing qualities of workers are high, the probabilities of their selection are also high, and this ensures the selection of high-quality workers. Because the number of workers in the network is vast, the platform explores the qualities of some workers by interacting with them, while the qualities of a much larger number of workers remain unknown [4], [12], [37]. Thus, this method actually selects only those workers whose qualities are already known. In the case of many unknown workers, workers with higher qualities may not be selected, which leads to local optimization. To solve this problem, in the UCB design, if a worker is selected a lot, his probability of being selected decreases, and vice versa, his probability of being selected increases, thus allowing the platform to explore the unknown workers. The MAB is composed of two actions, exploration and exploitation [4], [12], [34]. Exploration is the trial selection of unknown workers in order to discover high-quality workers, while exploitation is the use of acquired information to select high-quality workers [4], [12], [37]. One of the most important assumptions for the MAB mechanism to work is that the platform will know the qualities of the workers' data once it has received it. In fact, this does not hold true for the actual MCS. Numerous studies on truth data discovery have shown that after the platform receives data from workers, it is very difficult to verify the authenticity of the data submitted by workers and determine the data qualities.

(3) The third type of study abandon the assumption that the platform knows the qualities of the data once obtaining the workers' data. The platform must adopt challenging strategies to obtain real truth, known as the Information

Elicitation Without Verification (IEWV) problem [45]. In practice, to ensure data quality, the first thing is to ensure that the data itself is truthful, which is the cornerstone of data quality. If the data is not truthful, the quality undoubtedly is very low [13], [51], [57]. In most cases, data truthfulness is often one of the most critical data quality issues. If the data sensed by the worker is consistent with the objective physical world, then its quality is high. Thus, for many MCS applications, truth data is high-quality data. Some truth data discovery methods have been proposed [13], [39], [47], such as the Mean method [40], the Weighted Mean method [47], the Median method [40], the Major Vote method [40], [47]. This approach is based on the assumption that most workers in MCS are trustworthy. Thus, if $k$ workers are employed to collect data for a monitoring object, the data provided by these $k$ workers are averaged [40], weighted average [47], and median [40] to estimate the true value. The idea of the weighted mean method [47] is that the $k$ data obtained obey a normal distribution, and based on the assumption that most workers are trustworthy workers, the data at the center of the normal distribution are closer to the true value and are thus given a large weight. In contrast, the data far from the normal distribution center are given a smaller weight so that the value obtained by further weighting is the estimated true value [47]. Such a method of obtaining the true value still needs to be improved. Firstly, although in the actual MCS, most workers are trustworthy, in the specific data collection, malicious workers tend to launch attacks in concert. When the proportion of malicious workers among the $k$ workers is large, the obtained values will have significantly deviated. Even worse, the data obtained will be completely false if the malicious workers are in the overwhelming majority among the $k$ workers. Unfortunately, malicious workers are more active and can easily dominate in a coordinated attack during local data collection. Secondly, the cost of this method is high. For one task, $k$ recruited workers mean $k$ times larger cost, which makes it challenging to apply this method in practice, although it is meaningful in theoretical research. The optimization of worker selection is lacking in these studies, which mainly employ multiple workers to collect data simultaneously and perform some kind of calculation to estimate the truth data. In such an approach, there is no need to select workers before data collection but to perform the calculation to obtain truth data after data is obtained. This is a shortcoming of this approach, suggesting that it needs to achieve global optimization.

In fact, the truth data evaluation above is a direct method of evaluating the qualities of data. The advantage of this method is that there is no need to evaluate the submitted data in advance but to obtain the truth data from these data afterward [40], [47]. However, this method is susceptible to interference from the data source, and obtaining high-quality data is difficult if the source data received is of poor quality. Another direct method of data quality evaluation is to directly obtain real data to check the quality of data submitted by workers. However, direct access to real data requires using a trusted sensing device specifically for data sensing as Ground Truth Data (GTD) at the same time and place as the workers. It is clear that this approach is difficult to adopt due to the time and cost, which have been systematically discussed in [49]-[51]. In most of the above approaches, an indirect evaluation method is used, i.e., the majority of researchers use worker quality as an indication of data quality [14]. According to the core principle of Gestalt Psychology, people's cognition, experience, and action have integrity and consistency [64]. As a result, from a statistical standpoint, the qualities of data represented by worker quality are, to some extent, acceptable [14]. Among them, the Degree Of Trust (DOT) is a frequently used indirect evaluation method [53], [56], [57]. The basic idea of the trust-based evaluation method is that a claim is likely to be true if it is provided by trustworthy sources (especially if by many of them) and a source is trustworthy if most of its claims are true. Based on this idea, most methods attempted to assign a larger weight to reliable sources as they are more important when inferring the truth [9]. Thus, one of the big problems in such a study is how to obtain workers' DOT, which is an issue of equal difficulty to the data quality evaluation [6]. The DOT is calculated in several ways as follows.

The calculation of DOT has been studied for a long time, and a set of calculation methods has been developed [53], [56], [57]. DOT calculation has several important concepts: direct, indirect, and combined trust [56]. Direct trust means that after a direct interaction between two parties (e.g., A and B), they can evaluate each other's behavior in this interaction and thus derive DOT of each other [56]. If A considered that the behavior of B achieved a good interaction outcome, then A will rate B with a high DOT. Conversely, if A considered B's behavior malicious, a low DOT will be given to B. Multiple direct DOT are obtained after multiple direct interactions. In general, it is necessary to combine the direct DOT obtained from many interactions over time into one direct DOT. The typical approach is based on giving different weights to the trusts obtained at different times, with small weights for DOT far from the current time and large weights for DOT close to the current time. The indirect DOT means that A and B have not interacted with each other directly, but B has interacted with C directly, allowing B to recommend C to A by offering the DOT of C [56]. As a result, A can obtain the DOT of C indirectly, and the DOT obtained is called indirect DOT. Moreover, indirect trust can be carried out through a longer chain, forming a Chain Of Trust (COT). However, the longer the COT is, the less trustworthy its trust evaluation is. The indirect DOT is calculated by multiplying the DOT on the COT. If there are multiple COT, the indirect DOT of each COT should be combined. An evaluated subject often has both direct and indirect DOT. Therefore, it is necessary to calculate them together to obtain a comprehensive evaluation [56]. The above trust evaluation methods have been widely used in many applications, such as service computing, product evaluation, and so on. However, these traditional trust evaluation methods are difficult to be applied to MCS for their variable and complex characteristics. Firstly, in MCS, there is only direct interaction between workers and the platform, but no interaction between workers. Therefore, the indirect DOT does not exist, and the only direct DOT is also very different from previous studies and difficult to obtain. As mentioned above, to obtain a direct trust relationship, both parties must be able to know the quality of the interaction. Nevertheless, in MCS, it is challenging for the platform to obtain the qualities of the submitted data by workers directly, so the platform cannot evaluate the workers directly. Since there is neither direct nor indirect trust, a more advanced approach is urgently needed in MCS to gain the DOT of workers.

There are already some methods to obtain workers' DOT in MCS. Since workers' DOT tends to characterize the qualities of their submitted data, workers' DOT is consistent with the qualities of workers' submitted data in many previous studies. For example, in vehicle MCS, some sensing devices deployed in the network collect data from mobile vehicles passing by. However, the vehicles also have some low DOT or maliciously report false to get a reward. This untrustworthy behavior of vehicles is similar to the behavior of workers in this paper. It is not easy to verify whether the data submitted by them are true [55]. Guo et al. [55] proposed a method in which collecting part of the data act as GTD from the sensing device by sending Unmanned Aerial Vehicles (UAVs) to verify the data submitted by the vehicles. If the data submitted by the vehicle is consistent with GTD, its DOT will increase, but if not, its DOT will decrease. This approach is relatively accurate in terms of the DOT due to the availability of comparable GTD [55]. However, this study only investigated how to evaluate the DOT of vehicles, but how to select from the global scope and evaluate the data collectors was still unsolved.

Finally, another important aspect is the incentive mechanism, where the auction is generally employed [4], [36], [63]. In auctions of MCS, the platform will first show the tasks to the public, and workers will choose the tasks they are interested in and report their bids. Eventually, the platform selects the optimal worker to complete the task based on the qualities and bids of the workers [4], [36], [63]. In this process, workers have the following characteristics: **(a) Selfishness**: Workers are reluctant to provide sensing data without compensation [63]. **(b) Individual Rationality**: The earnings for each worker must be non-negative, which means that once a worker participates in the task, his reward from the platform must be higher than the cost of sensing the data [63]. **(c) Untruthfulness**: Since performing sensing tasks can incur some resource consumption or lead to a privacy breach, workers may intentionally submit low-quality or even fake sensing data to get a large reward at a minimum cost [63]. **(d) Uncertainty** [63]: The sensing abilities of workers depends on the sensing devices and subjective attitudes, etc. For example, temperature measurement tasks' results depend on both workers' temperature measurement devices and their subjective intentions [65]. From the above characteristics of workers, we can see that the truth of data, data quality, and bid are the issues to take into account in the auction mechanism, which previous studies tend to ignore some of the elements. From the platform perspective, it is difficult to measure the platform's benefits because large applications are performed by many workers completing multiple small tasks in many MSC applications. As a result, in some studies, the qualities of the data given by workers are used to assess the platform's benefits. The underlying idea is that when the qualities of the data submitted by workers are higher, the qualities of the constructed applications will be higher, and thus the benefit of the platform will be greater [4], [12]. In many studies, it is a logical and typical approach to evaluate the platform's benefits in terms of the qualities of the submitted data. In this approach, workers initiate their bid to the platform, and the platform will select the workers with the high qualities and low bids after it receives all the bids [4], [12]. In such an auction, the platform often has a specific budget and cannot exceed it. Overall, it is challenging to design an auction mechanism that meets several expectations, such as budget feasibility, truthfulness, individual rationality, and computational efficiency [63].

## 3. System model and problem statement

### 3.1 System Overview

We consider an MCS system consisting of the platform, some requesters, and a crowd of unknown workers. The requester wants to recruit some workers to perform data collection tasks (e.g., collecting data on traffic flow, water quality, air quality, etc.) in urban areas periodically within a limited budget $B$. The tasks and unknown workers are defined as follows:

**Definition 1 (Task, Weight, Round).** The requester publishes $\mathcal{M}$ location-sensitive data collection tasks via the platform, denoted by $\mathcal{D} = \{d_1, d_2 \dots, d_\mathcal{M}\}$. Each task $d_j$ is attached with a weight $w_j$ to indicate its importance. Moreover, the data collection is divided into multiple rounds, denoted by $t \in \{1, 2, \dots\}$.

**Definition 2 (Unknown Workers, Sensing Rate, Cost).** There are $\mathcal{N}$ unknown workers, denoted by $\mathcal{S} = \{s_1, s_2 \dots, s_\mathcal{N}\}$. We let $v_{j,t}^i$ represents the reported data by worker $s_i$ for task $d_j$ in the $t$-th round. Due to scheduling conflicts or malicious behaviors, workers will not always honestly complete their assigned work on time, resulting in them reporting false data in an attempt to deceive the platform. So, we let $r_{i,t} \in [0,1]$ denote the Sensing Rate (SR) of worker $s_i$ completing tasks in the $t$-th round. Each $r_{i,t}$ follows an unknown distribution with an unknown expectation $r_i$. Each worker $s_i$ can perform a set of preferred tasks in each round $t$, denoted by $\mathcal{D}_{i,t}$, with a certain cost $c_{i,t}$ for the whole tasks set. We assume that the cost for a worker to complete each task will not exceed the range $[c_{min}, c_{max}]$.

In Def. 2, the $r_{i,t}$ is determined by some worker side factors, such as sensing ability, personal willingness, personal emergency and so on, which bring about the variance of actual SRs for different round. Hence, for $\forall$ round $t' \neq t$, $r_{i,t'}$ may not be equal to $r_{i,t}$. SRs directly reflect a worker's credit and can be considered as the most basic sensing quality. Because if a worker does not even complete the task and his reported data is fake, then there is no sensing quality.

The MCS system uses a reverse auction to recruit unknown workers. The platform, the unknown workers, and the requesters are considered auctioneers, workers, and

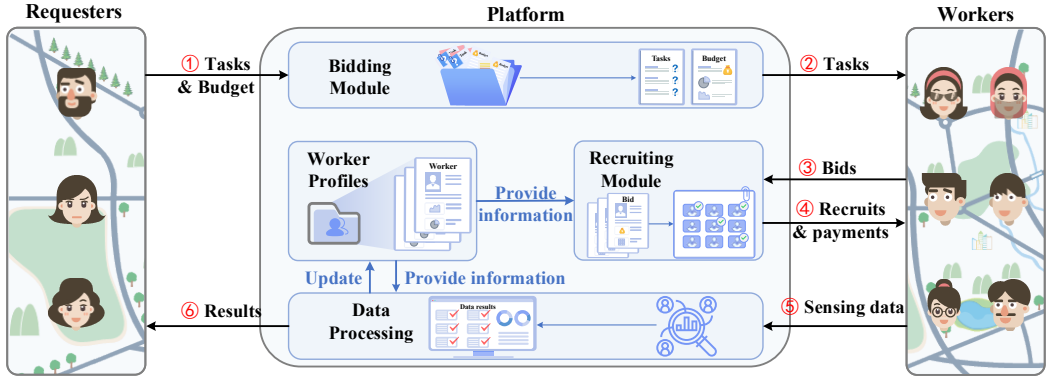

**Fig. 1: Detailed workflow in MCS.**

buyers, respectively. The goods for sale are services to perform data collection tasks. First, workers submit bids for the tasks they prefer. Then, the platform determines the auction winner (i.e., makes the worker recruitment decision) and calculates the corresponding payment, which is defined as follows:

**Definition 3 (Bid).** In each round $t$, each worker $s_i$ submits a task-bid pair $\beta_i = \langle \mathcal{D}_{i,t}, b_{i,t} \rangle$ to the platform, where $b_{i,t}$ is the claimed cost for completing all tasks in $\mathcal{D}_{i,t}$. Since the true sensing cost $c_{i,t}$ is a piece of private information, the strategic worker $s_i$ might misreport its cost to obtain higher utility. That is, the claimed cost $b_{i,t}$ is not necessarily equal to the true cost $c_{i,t}$. Additionally, we denote all workers' bidding profiles as $\beta = \{\beta_1, \beta_2, \cdots, \beta_t, \cdots\}$, in which $\beta_t = \{\beta_{1,t}, \beta_{2,t}, \cdots, \beta_{\mathcal{N},t}\}$.

**Definition 4 (Payment).** In each round $t$, the platform will determine a payment for the worker $s_i$ who wins the auction, denoted as $p_{i,t}$. All workers' payment profiles are represented by $\mathcal{P} = \{\mathcal{P}_1, \mathcal{P}_2, \cdots, \mathcal{P}_t, \cdots\}$, in which $\mathcal{P}_t = \{p_{1,t}, p_{2,t}, \cdots, p_{\mathcal{N},t}\}$.

Based on the reverse auction model, the detailed workflow of worker recruitment is shown in Fig. 1. First, the requester posts all the sensing tasks and the total budget on the platform. Next, all workers submit their task bid pairs to the platform. Then, recruiting workers proceeds round by round until the budget is exhausted. In each round, the platform first conducts exploitation to determine the auction winner (i.e., selects the workers to be recruited) and calculates the corresponding payments. Then, the platform will inform the recruited workers to perform their preferred tasks. After completing the tasks, the workers submit their sensing data to the requester via the platform. Since the data reported by the workers may be fake for their unknown SRs, the platform needs to explore their SRs according to the submitted data. However, the information available to the platform is only the workers' reported data, so the platform often needs to use some supervision-based methods to infer whether the workers' reported data are real or fake.

**Definition 5 (Supervised SR Learning, GTD).** The platform needs to obtain some posterior Ground Truth Data (GTD) for comparison to evaluate workers' authenticity, especially at the beginning stage of MCS, where workers' SRs are entirely unknown. Specifically, as a service provider, the platform can maintain several employees to collect data synchronously with workers for the same tasks. The data sensed by the employees can be regarded as GTD. And we let $v_{j,t}^{GTD}$ represent the GTD of the task $d_j$ in the $t$-th round. Thus, the platform can verify the workers' submissions by comparing their reported data with the GTD. Since this process requires extra employees to collect GTD, supervised SR learning is named.

**Definition 6 (Self-Supervised SR Learning, ETD).** In the real MCS system, the number of employees specializing in collecting GTD is often minimal. It is impossible to rely entirely on the GTD provided by the employees to verify the workers' data. Therefore, the platform needs to obtain the Estimated Truth Data (ETD) as a substitute based on the data submitted by multiple workers themselves, which will be described in detail in Section 4. We can then use the ETD to verify the workers' data in a GTD-like way. Since this method only requires the data submitted by workers, it is named self-supervised SR learning.

## 3.2 CMAB Modeling and Problem Formulation

With all workers' SRs initially unknown, how to identify auction winners and calculate the corresponding payments is the most critical issue for the platform in the whole worker recruitment process.

For one thing, the platform needs to learn about the value of workers' SRs (so-called exploration) for better recruitment performance; For another, the learned SR knowledge will be used to make the best winner decision (so-called exploitation). It is an online learning and decision-making process. Multi-Armed Bandit (MAB) is a widely used reinforcement learning model for online decision-making in uncertain conditions. A MAB model consists of a slot machine with multiple arms, each arm being associated with a reward taken from an unknown distribution. The player needs to pull some arms round by round according to a bandit policy to maximize the cumulative reward. In this paper, we model unknown worker recruitment as a novel $K$-arm CMAB decision process, defined as follows:

**Definition 7 (CMAB Modeling).** The auction winner selection is unknown worker recruitment and can be modeled as $K$-armed CMAB. The platform is a player, each worker in $\mathcal{S}$ is an arm, and the SR of each recruited worker is the reward of pulling the corresponding arm. In each round, the platform will pull $K$ arms. Pulling the $i$-th arm worker

TABLE 1: Description of major notations

| Notations | Description |
|---|---|
| $\mathcal{M}, \mathcal{N}$ | Number of sensing tasks and workers, respectively |
| $\mathcal{D}, \mathcal{S}$ | Sets of sensing tasks and workers, respectively |
| $\mathcal{D}_{i,t}$ | Sets of preferred tasks for $s_i$ in round $t$ |
| $w_j$ | Weight of the task $d_j$ |
| $r_{i,t}$ | Sensing rate of $s_i$ in round $t$ |
| $r_i$ | Expectation of the distribution followed by $r_{i,t}$ |
| $\hat{r}_{i,t}$ | Sample mean of $s_i$'s sensing rate until round $t$ |
| $\hat{r}_t^+$ | UCB index of $s_i$ |
| $c_{i,t}$ | Real cost for completing all tasks in $\mathcal{D}_{i,t}$ |
| $b_{i,t}$ | Claimed cost for completing all tasks in $\mathcal{D}_{i,t}$ |
| $p_{i,t}$ | Payment to $s_i$ in round $t$ |
| $v_{j,t}^i$ | Reported data by $s_i$ for $d_j$ in round $t$ |
| $v_{j,t}^{GTD}$ | GTD for $d_j$ in round $t$ |
| $\hat{u}_{j,t}^{1st}$ | First ETD for $d_j$ in round $t$ |
| $\hat{u}_{j,t}^{2nd}$ | Second ETD for $d_j$ in round $t$ |
| $\varphi_{i,t}$ | Whether $s_i$ is selected in round $t$ |
| $\xi_{i,j}^t$ | Whether $s_i$ truthfully finish $d_j$ in round $t$ |
| $n_{i,t}$ | Recruited times of $s_i$ until round $t$ |
| $\rho_i$ | Revenue-cost-ratio of worker $s_i$ |

$s_i$ is recruited.

Under the CMAB model, the worker recruitment problem is turned into a budget-feasible bandit policy determination problem to maximize the total expected revenue. The bandit policy and revenue are defined as follows:

**Definition 8 (Budget-Feasible Bandit Policy).** A bandit policy $\phi$ is a sequence of maps: $\{\phi_1, \phi_2, \cdots, \phi_t, \cdots\}$, where $\phi_t = \varphi_{1,t}, \varphi_{2,t}, \cdots, \varphi_{\mathcal{N},t}$. $\varphi_{i,t} \in \{0,1\}$. Moreover, the total cost of worker recruitment is no larger than the budget $B$.

**Definition 9 (Revenue).** Under the CMAB model, the total revenue refers to the recruited workers' total weighted number of effective sensing. Effective sensing is counted if a worker completes the task and reports accurate data. On the contrary, if the worker does not complete the task or gives fake data, the recruitment is considered to gain no revenue, although the money has been spent. We let $\xi_{i,j}^t \in \{0,1\}$ indicate whether the worker $s_i$ truthfully finish the task $d_j$ in the $t$-th round, so the revenue can be defined as

$$\mathrm{R}(\phi) = \sum_t \sum_{i=1}^{\mathcal{N}} \sum_{j|d_j \in \mathcal{D}_{i,t}} \omega_j \xi_{i,j}^t \varphi_{i,t}. \quad (1)$$

Our objective is to maximize the total expected revenue under the budget constraint. Thus, the winner selection problem can be formulated as follows:

*Maximize:*

$$E[R(\phi)] \quad (2)$$

*Subject to:*

$$\sum_t \sum_{i=1}^{\mathcal{N}} p_{i,t} \leq B \quad (3)$$

$$\sum_{i=1}^{\mathcal{N}} \varphi_{i,t} = K, \forall t \quad (4)$$

$$\varphi_{i,t} \in \{0,1\}, \forall i | s_i \in \mathcal{S}, \forall t \quad (5)$$

Here, Eq. (3) ensures that the total payment is within the budget. Eq. (4) indicates that $K$ workers are recruited in each round. Eq. (5) implies that all worker recruitment decisions are binary.

In addition, the platform needs to determine the payment for the auction winner. To ensure that workers are willing to perform tasks and truthfully report their costs, payment calculations should satisfy both truthfulness and personal rationality. Before this, we first define the concept of utility:

**Definition 10 (Worker's Utility).** The utility of worker $s_i$ in the $t$-th round is the payment minus cost:

$$u_{i,t} = (p_{i,t} - c_{i,t})\varphi_{i,t}. \quad (6)$$

Here, $u_{i,t}$ and $p_{i,t}$ actually are the functions of bid $b_i^t$, i.e., $u_{i,t} = u_{i,t}(b_{i,t})$, and $p_{i,t} = p_{i,t}(b_{i,t})$.

Then, our mechanism is proven to satisfy the properties of truthfulness, individual rationality and computational efficiency.

**Definition 11 (Truthfulness).** Let $b_{i,t}$ be an arbitrary bid of worker $s_i$ to complete tasks in $\mathcal{D}_{i,t}$, and $u_{i,t}(b_{i,t})$ is the utility obtained by worker $s_i$ in the $t$-th round. Then, if

$$u_{i,t}(b_{i,t}) \leq u_{i,t}(c_{i,t}), \forall i, \forall t. \quad (7)$$

we say that the mechanism is truthful.

**Definition 12 (Individual Rationality).** For $\forall i, st. s_i \in N$, if $s_i$'s utility is non-negative,

$$u_{i,t} \geq 0, \forall i, \forall t. \quad (8)$$

we say that the mechanism is individual rationality.

**Definition 13 (Computational Efficiency).** A mechanism is computationally efficient if it generates the results and terminates in polynomial time.

For ease of reference, we have classified and summarized the significant notations in this paper in Table 1.

## 4. The SCMABA mechanism

In this section, we present the incentives Semi-supervision based Combinatorial Multi-Armed Bandit reverse Auction (SCMABA), whose process is demonstrated in Fig. 2. We first introduce the basic idea of SCMABA, then present the detailed algorithm, followed by an illustrative example to show how our SCMABA mechanism works.

### 4.1 Basic Idea

Since the whole worker recruitment is an online learning and decision-making process, we separate it into the exploration and exploitation phases under the CMAB model.

Accordingly, the budget is also divided into two phases. The first one is the exploration phase, where we uniformly select the workers to execute tasks and pay each recruited worker the maximum cost he might incur. In the second phase, named exploitation, all of the remaining budgets are used to maximize the total expected revenue based on the acquired empirical knowledge. We adopt a UCB-based greedy strategy to select workers and compute payments based on the second price scheme to guarantee truthfulness. In what follows, the methods adopted in each phase are discussed separately.

Since supervised and self-supervised SR learning respect-tively used in the two phases, the whole process is considered semi-supervised.

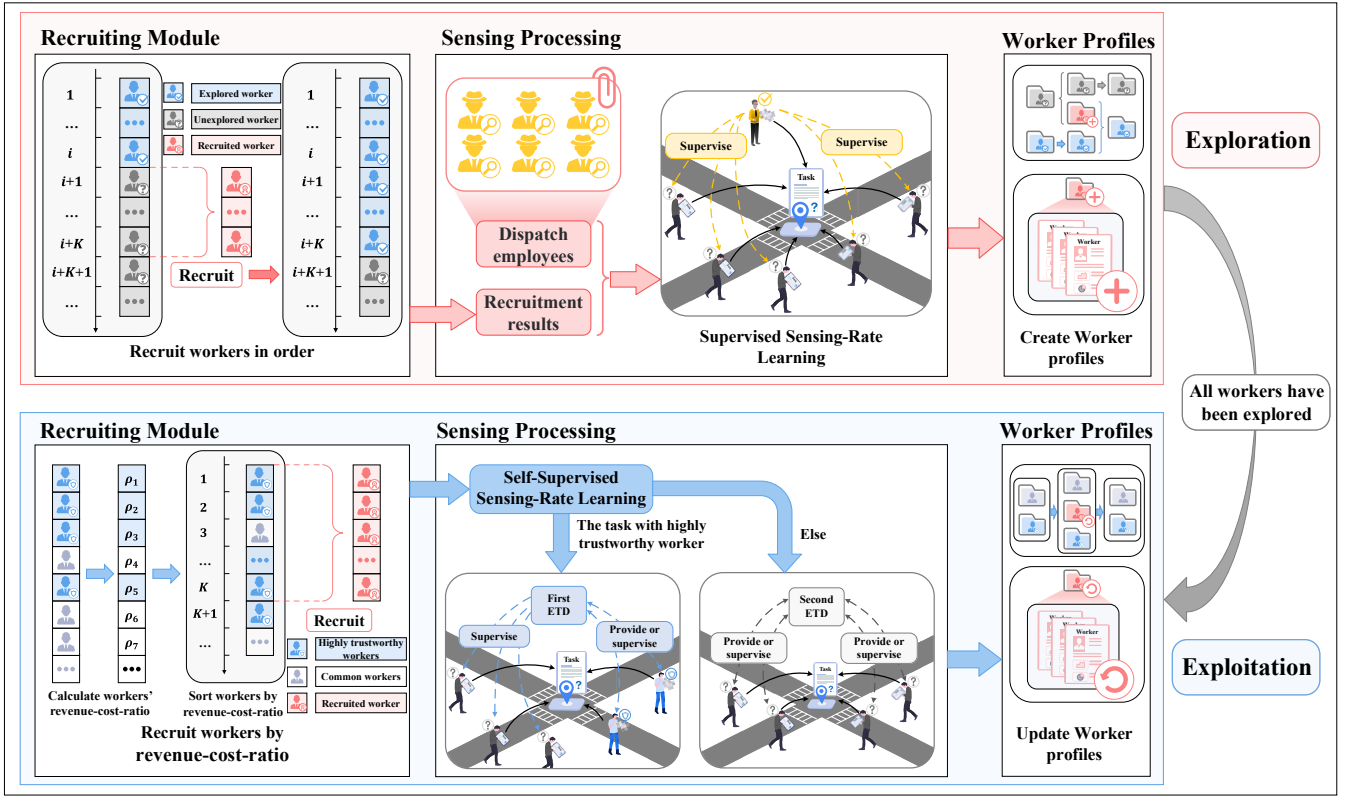

**Fig. 2: The framework of SCMABA:** SCMABA consists of two phases, exploration and exploitation. During the exploration phase, since the quality of the workers is unknown, the mechanism needs to perform an initial assessment of the workers, meaning that the mechanism will recruit all workers at least once and give them an expected SR. Specifically, in the recruitment model, the mechanism first treats workers equally and recruits them in order. Then, in the sensing process, we conduct the sensing process by supervised SR learning, i.e., we simultaneously dispatch employees in parallel to obtain accurate judgments to supervise workers. Finally, we create the worker profiles of the recruited workers based on the learned SR. During the exploitation phase, we will consider multiple workers' information learned to recruit the best workers in our opinion in order to gain greater profitability. In the recruitment model, we calculate the revenue-cost-ratio of all workers based on data such as the worker's offer and the worker's SR, and recruit the worker with the highest revenue-cost-ratio. In the perception process, for tasks with highly trustworthy workers, the data of highly trustworthy workers are used to obtain the first ETD as the judging criterion; for other tasks, the data submitted by all workers are used to obtain the second ETD as the judging criterion. In the worker profiles, the information and records of these recruited workers are updated.

### 4.1.1 Exploration Phase by Supervised SR Learning

In this phase, the platform tentatively recruits workers to perform sensing tasks to learn their expected SRs. Since the SRs of each worker is unknown in advance, we need to treat each worker equally. Given a budget $B$, the platform needs to execute an initial supervised SR learning for all workers on a rotating basis. That is, workers $\{s_1, s_2, \cdots, s_K\}$ will be recruited in the first round, workers $\{s_{K+1}, s_{K+2}, \cdots, s_{2K}\}$ will be recruited in the second round, and so on, until all workers have been learned at least once.

We keep the two vectors $n_t = \{n_{1,t}, n_{2,t}, \cdots, n_{\mathcal{N},t}\}$ and $\hat{r}_t = \{\hat{r}_{1,t}, \hat{r}_{2,t}, \cdots, \hat{r}_{\mathcal{N},t}\}$ as empirical knowledge learned from long-term history. More specifically, $n_{i,t}$ is the number of tasks that $s_i$ will complete at the end of the $t$-th round, and $\hat{r}_{i,t}$ is the sample mean of SR for worker $s_i$ by then.

The scenario of supervised SR learning is demonstrated in Fig. 3. Taking traffic monitoring as an example, different workers monitor traffic flow data at different intersections.

In supervised SR learning, the platform will dispatch employees to collect data on the same tasks in parallel with the workers to supervise the workers. Specifically, employees are members within the platform who are trusted and the data they submit will be considered GTD. The employees and workers measure the traffic data simultaneously and report the data to the platform. The platform supervises the workers by comparing the data collected by employees and workers. Finally, the platform creates worker profiles for these workers based on this data.

Assuming that $\varepsilon_1$ is a very small positive number, if $\left| v_{j,t}^i - v_{j,t}^{GTD} \right| < \varepsilon_1$ is satisfied, the data sensed by $s_i$ for $d_j$ in the $t$-th round will be considered truthful. And we let $\hat{\xi}_{i,j}^t \in \{0,1\}$ denote our judgment on whether worker $s_i$ truthfully finish the task $d_j$ in the $t$-th round. After that, the workers' SRs are learned and recorded, and the worker profiles will be updated. In our setting, once a worker $s_i$ is recruited in a round, the corresponding quality would be learned $|\mathcal{D}_{i,t}|$ times. Thus, $n_{i,t}$ will be updated as follows:

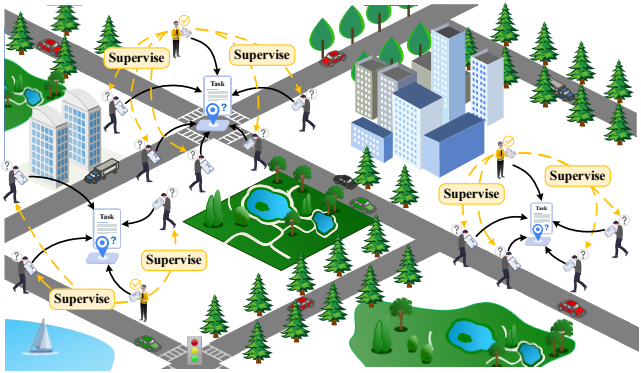

**Fig. 3: Supervised SR learning for exploration.**

$$n_{i,t} = \begin{cases} n_{i,t-1} + |\mathcal{D}_{i,t}|, \varphi_{i,t} = 1; \\ n_{i,t-1}, \varphi_{i,t} = 0. \end{cases} \quad (9)$$

The sample mean of worker $s_i$'s SR is updated as

$$\hat{r}_{i,t} = \begin{cases} \dfrac{\hat{r}_{i,t-1}n_{i,t-1} + \sum_{j|d_j \in \mathcal{D}_{i,t}} \omega_j \xi_{i,j}^t}{n_{i,t-1} + |\mathcal{D}_{i,t}|}, \varphi_{i,t} = 1; \\ \hat{r}_{i,t-1}, \varphi_{i,t} = 0. \end{cases} \quad (10)$$

The SRs learned in exploration will be used for worker recruitment in the exploitation phase. Here, we define UCB indexes for workers rather than using the sample means directly, denoted by $\hat{r}_t^+ = \{\hat{r}_{1,t}^+, \hat{r}_{2,t}^+, \cdots, \hat{r}_{\mathcal{N},t}^+\}$, where we take the uncertainty of estimation into consideration. That is, the estimated sample mean may have a certain error, and the additive factor $\varepsilon_{i,t}$ makes the less selected worker have more chances to be selected. The UCB index $\hat{r}_{i,t}^+$ of worker $s_i$ is

$$\hat{r}_{i,t}^+ = \hat{r}_{i,t} + \mathcal{U}_{i,t}. \quad (11)$$

$$\mathcal{U}_{i,t} = \sqrt{\frac{\delta \cdot \ln\left(\sum_{i'|s_{i'} \in \mathcal{S}} n_{i',t}\right)}{n_{i,t}}}. \quad (12)$$

where $\mathcal{U}_{i,t}$ is a factor for exploration, when a worker $s_i$ has not been selected for a long time, his $\mathcal{U}_{i,t}$ will increase, thus increasing the probability of his selection. And $\delta$ is a positive hyperparameter that brings flexibility to our policy. The exploration process terminates until all workers have been learned at least once due to the unknown workers' SRs. To guarantee truthfulness and individual rationality. We set a maximum cost for each recruited worker $s_i$ in the exploration phase, which might incur as the payment, i.e., $|\mathcal{D}_{i,t}|c_{max}$. The exploration process terminates until all workers have been learned at least once.

#### 4.1.2 Exploitation Phase by Self-supervised SR Learning

In the exploitation phase, we determine the winning workers and the corresponding payments based on the UCB indexes learned in the exploration phase. We assume that all workers report their costs truthfully, i.e., $b_i = c_i$, which is to be proved reasonable in Subsection 4.3. Since the worker recruitment problem in our paper can be regarded as a series of special 0-1 knapsack problems, we adopt a UCB-based greedy strategy to select workers. For each worker $s_i$, we define the revenue-cost-ratio $\rho$ as $\left(\sum_{j|d_j \in \mathcal{D}_{i,t}} \omega_j \hat{r}_{i,t}^+\right)/b_{i,t}$. The strategy for recruiting workers is to select workers with the highest $K$ $\rho$s in each round. Thus, we first calculate $\rho$ for each worker and then sort the workers in $\mathcal{S}$ into $\mathcal{A}_t = (a_{1,t}, a_{2,t}, \cdots, a_{\mathcal{N},t})$ such that $\rho_{a_{1,t}} \geq \rho_{a_{2,t}} \geq \cdots \geq \rho_{a_{\mathcal{N},t}}$. Then, we greedily select the best workers for the winning workers' set $\mathcal{A}_t'$ and compute the payments for them.

To guarantee the truthfulness, the platform should pay all winning workers for the critical payment, according to [66]. The critical payment is equal to the highest bid that still guarantees the bid win. More specifically, the critical payment of a winning worker $s_i$ should be calculated based on the bid of the $(K+1)-th$ worker in $\mathcal{S}$, i.e. $\frac{\sum_{j|d_j \in \mathcal{D}_{i,t}} \omega_j \hat{r}_{i,t}^+}{\sum_{j|d_j \in \mathcal{D}_{a_{K+1,t}}} \omega_j \hat{r}_{a_{K+1,t}}^+} b_{a_{K+1,t}}$. When the critical payment for a worker $s_i$ is larger than $|\mathcal{D}_i|c_{max}$, we set the payment as $|\mathcal{D}_i|c_{max}$. Hence, $\forall s_i \in \mathcal{S}$, the payment of worker $s_i$ in the exploitation phase is calculated as:

$$p_{i,t} = min\left\{\frac{\sum_{j|d_j \in \mathcal{D}_{i,t}} \omega_j \hat{r}_{i,t}^+}{\sum_{j|d_j \in \mathcal{D}_{a_{K+1,t}}} \omega_j \hat{r}_{a_{K+1,t}}^+} b_{a_{K+1,t}}, |\mathcal{D}_i|c_{max}\right\}. (13)$$

After each worker has been recruited and the corresponding quality has been learned at least once in the supervised phase for exploration, the platform will exploit the learned UCB indexes computed by Eqs. (11) and (12) to execute a reverse auction in the next round to determine $K$ winning workers based on each worker $s_i$'s $\rho$ (i.e., $\left(\sum_{j|d_j \in \mathcal{D}_{i,t}} \omega_j \hat{r}_{i,t}^+\right)/b_{i,t}$), and compute the corresponding payment for each winning worker based on the bid of the $(K+1)$-th worker and Eq. (13).

After each auction round in the exploitation phase, the platform will adopt the method of self-supervised SR learning to infer the recruited workers' current SRs and update their UCB based on the inferred SRs and Eqs. (9)-(12), which will be described in detail below.

Before describing self-supervised SR learning in the exploitation phase, we first define highly trustworthy workers: when a worker $s_i$'s recorded average SR $\hat{r}_{i,t}$ is greater than a threshold $\theta$, we consider him as a highly trustworthy worker in the $t$-th round. And we maintain a set $S_t^{HT}$ to contain all the highly trustworthy workers in the $t$-th round.

The scenario of self-supervised SR learning is shown in Fig. 4. Taking traffic monitoring as an example, different workers monitor traffic flow data at different intersections. First, the platform recruits workers, common workers and highly trustworthy worker, to collect traffic flow data. Second, the platform obtains the ETD for these tasks. For tasks with highly trusted workers, the platform obtains the first ETD based on the data provided by highly trustworthy workers; for other tasks, the platform obtains the second ETD based on the data reported by all workers. Then, the platform evaluates the data quality reported by workers based on the obtained ETD to supervise them. Finally, the platform updates the worker profiles based on the data obtained.

For a sensing task in the $t$-th round, if there is at least one highly trustworthy worker, then we consider that there is a first ETD for this task in the current round, denoted as $u_{j,t}^{1st}$.

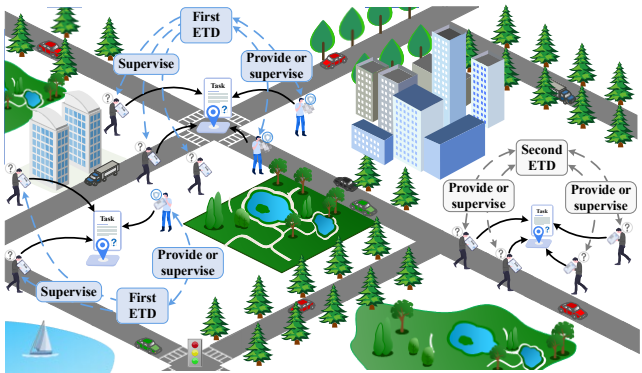

**Fig. 4: Self-supervised SR learning for exploitation.**

And the first ETD is the weighted average of the reported data of each highly trustworthy worker, where the weight is an authenticity term $\lambda_{j,t}^i$ that we introduce for each worker. $\lambda_{j,t}^i$ depicts the authenticity of data submitted by worker $s_i$ for task $d_j$ in the $t$-th round. We first initialize $\hat{\lambda}_{j,t}^i = \hat{r}_{i,t}$ and gradually improve the performance of estimation, we need to iterate continuously to update ETD and $\hat{\lambda}_{j,t}^i$. Specifically, when a worker's estimation authenticity $\hat{\lambda}_{j,t}^i$ is larger, then the impact of his or her reported data on the ETD will be more significant. Meanwhile, when a worker's reported data is closer to the current ETD, his or her estimation authenticity $\hat{\lambda}_{j,t}^i$ will be correspondingly larger. Eventually, we can obtain $u_{j,t}^{1st}$ by iterating the following two updates till convergence [67].

$$\hat{u}_{j,t}^{1st} = \frac{\sum_{i|d_j \in \mathcal{D}_{i,t} \& s_i \in S_t^{HT}} \hat{\lambda}_{j,t}^i v_{j,t}^i \varphi_{i,t}}{\sum_{i|d_j \in \mathcal{D}_{i,t} \& s_i \in S_t^{HT}} \hat{\lambda}_{j,t}^i \varphi_{i,t}}. \quad (14)$$

$$\frac{1}{\hat{\lambda}_{j,t}^i} = \frac{\left(v_{j,t}^i - \hat{u}_{j,t}^{1st}\right)^2 \varphi_{i,t}}{\sum_{i|d_j \in \mathcal{D}_{i,t} \& s_i \in S_t^{HT}} \varphi_{i,t}}. \quad (15)$$

For a task $d_j$ with first ETD, if $\left| v_{j,t}^i - u_{j,t}^{1st} \right| < \varepsilon_2$ is satisfied, where $\varepsilon_2$ is a minimal positive number, the data sensed by $s_i$ for $d_j$ in the $t$-th round are considered truthful.

For a task without GTD and first ETD, we can directly use the weighted average of all workers as the second ETD, denoted as $u_{j,t}^{2nd}$. The calculation of $u_{j,t}^{2nd}$ is basically the same as the principle of $u_{j,t}^{1st}$, except that the objects of calculation is changed from the highly trustworthy workers in $u_{j,t}^{1st}$ to all the workers recruited for the task $d_j$ in the $t$-th round. Therefore, in the absence of $u_{j,t}^{1st}$, we need to compute $u_{j,t}^{2nd}$ by iterating the following two updates till convergence [67].

$$\hat{u}_{j,t}^{2nd} = \frac{\sum_{i|d_j \in \mathcal{D}_{i,t}} \hat{\lambda}_{j,t}^i v_{j,t}^i \varphi_{i,t}}{\sum_{i|d_j \in \mathcal{D}_{i,t}} \hat{\lambda}_{j,t}^i \varphi_{i,t}}. \quad (16)$$

$$\frac{1}{\hat{\lambda}_{j,t}^i} = \frac{\left(v_{j,t}^i - \hat{u}_{j,t}^{2nd}\right)^2 \varphi_{i,t}}{\sum_{i|d_j \in \mathcal{D}_{i,t}} \varphi_{i,t}}. \quad (17)$$

And if $\left| v_{j,t}^i - u_{j,t}^{2nd} \right| < \varepsilon_3$ is satisfied, where $\varepsilon_3$ is also a small positive number, the data sensed by $s_i$ for $d_j$ in the $t$-th round are considered truthful.

---

**Algorithm 1**: *The SCMABA Mechanism*

**Input:** $\mathcal{D}$, $\mathcal{S}$, $B$, $K$, $c_{max}$, $\{\omega_j | d_j \in \mathcal{D}\}$, $\beta$, $\delta$, $v_{j,t}^i$, $\tau$, $\varepsilon_1$, $\varepsilon_2$, $\varepsilon_3$
**Output:** $\phi$, $\mathcal{P}$

1: Initialize $t = 0$, $\varphi_{i,t} = 0$, $p_{i,t} = 0$, $\xi_{i,j}^t = 0$, $\forall s_i \ \forall t$;
2: $B_t = B$; // Exploration by Supervision
3: **while** $t \leq \frac{N}{K}$ **do**
4:    $t = t + 1$; $\mathcal{K} = \emptyset$;
5:    **for** $a = 1 : K$ **do**
6:       $i = \{[(t-1)K + a - 1] \ mod \ N\} + 1$;
7:       $\mathcal{K} = \mathcal{K} \cup s_i$;
8:    **if** $\sum_{i|s_i \in \mathcal{K}} |\mathcal{D}_{i,t}| c_{max} \leq B_{t-1}$ **then**
9:       **foreach** $i \in \{i | s_i \in \mathcal{K}\}$ **do**
10:          $\varphi_{i,t} = 1$; $p_{i,t} = |\mathcal{D}_{i,t}| c_{max}$;
11:          **foreach** $j \in \{j | d_j \in \mathcal{D}_{i,t}\}$ **do**
12:             Dispatch the employees to collect $v_{j,t}^{GTD}$;
13:             **if** $|v_{j,t}^i - v_{j,t}^{GTD}| < \varepsilon_1$ **then** $\xi_{i,j}^t = 1$;
14:       **foreach** $i \in \{i | s_i \in \mathcal{S}\}$ **do**
15:          Update $n_{i,t}$, $\hat{r}_{i,t}$, $\hat{r}_{i,t}^+$;
16:       Update $B_t = B_{t-1} - \sum_{i|s_i \in \mathcal{K}} p_{i,t}$;
17:    **else break**;
18: Update $S_t^{HT}$;
19: **While** $t++$ **do** // Exploitation by self-supervision
20:    **foreach** $i \in \{i | s_i \in \mathcal{S}\}$ **do**
21:       Calculate the $\rho$ by $\frac{\sum_{j|d_j \in \mathcal{D}_{i,t}} \omega_j \hat{r}_{i,t}^+}{b_{i,t}}$;
22:    Sort the workers $\mathcal{S}$ into $\mathcal{A}_t$: $\rho_{a_{1,t}} \geq \cdots \geq \rho_{a_{N,t}}$;
23:    Select the best $K$ workers: $\mathcal{A}_t' = (a_{1,t}, a_{2,t}, \cdots, a_{K,t})$;
24:    **foreach** $i \in \{i | s_i \in \mathcal{A}_t'\}$
25:       Calculate $p_{i,t}$ by Eq. (13);
26:    **if** $\sum_{i|s_i \in \mathcal{A}_t'} p_{i,t} \leq B_{t-1}$ **then**
27:       **foreach** $i \in \{i | s_i \in \mathcal{A}_t'\}$ **do** $\varphi_{i,t} = 1$;
28:       Update $B_t = B_{t-1} - \sum_{i|s_i \in \mathcal{A}_t'} p_{i,t}$;
29:       **foreach** $j \in \{j | d_j \in \mathcal{D}\}$ **do**
30:          **if** exists $i \in \{i | d_j \in \mathcal{D}_{i,t}$ and $s_i \in S_t^{HT}\}$ **do**
31:             Calculate $u_{j,t}^{1st}$ by Eqs. (14) and (15);
32:             **if** $|v_{j,t}^i - u_{j,t}^{1st}| < \varepsilon_2$ **then** $\xi_{i,j}^t = 1$;
33:          **else** Calculate $u_{j,t}^{2nd}$ by Eqs. (16) and (17);
34:             **if** $|v_{j,t}^i - u_{j,t}^{2nd}| < \varepsilon_3$ **then** $\xi_{i,j}^t = 1$;
35:       **foreach** $i \in \{i | s_i \in \mathcal{S}\}$ **do**
36:          Update $n_{i,t}$, $\hat{r}_{i,t}$, $\hat{r}_{i,t}^+$;
37:       Update $S_t^{HT}$;
38:    **else break**;
39: **return** $(\phi, \mathcal{P})$;

---

Due to the decreasing accuracy of GTD, first ETD and second ETD inferences, the stringency of verify for the three conditions are supposed to be differentiated. Specifically, it should satisfy $\varepsilon_1 < \varepsilon_2 < \varepsilon_3$, Thus, as far as possible, the truthful data reported by workers will not be misjudged as falsified data.

In the following rounds, the platform will similarly in turn conduct the reverse auction and the quality learning until the total budget is exhausted.

### 4.2 The Detailed Algorithm

According to the above solution, the proposed SCMABA mechanism is depicted in Algorithm 1. In the beginning, we initialize all $t = 0$, $\varphi_{i,t} = 0$, $p_{i,t} = 0$, $\xi_{i,j}^t = 0$ (Step 1). Then, the initial exploration phase begins. We first tentatively recruit each worker to complete tasks and dispatch the employees to get the GTD in order to learn and estimate workers' SRs by supervision (Steps 3-12). For each recruited worker, the algorithm verifies his submission by comparing the reported data with GTD (Step 13). If $\left| v_{j,t}^i - v_{j,t}^{GTD} \right| < \varepsilon_1$ is satisfied, the data sensed by $s_i$ for $d_j$ in the $t$-th round will be considered truthful. Finally, $n_{i,t}$, $\hat{r}_{i,t}$, $\hat{r}_{i,t}^+$ of will each worker be updated by Eqs. (9)-(12) (Step 15). The initial exploration phase terminates until each worker has been selected at least once. After the initial exploration phase terminates, the algorithm will record the set of highly trustworthy workers $S_t^{HT}$ (Step 18).

We have preliminarily learned the SR information of each worker in the initial exploration phase. Then in the next round, we can use the SR information to select workers and compute payments (i.e., exploitation). In Steps 21-22, we first calculate $\rho$ for each worker and then sort the workers in a non-increasing order of their $\rho$s. Then, we greedily select the best $K$ workers into a winning set $\mathcal{A}_t'$ and determine the corresponding payments according to Eq. (13) (Steps 23-25). Next, the algorithm checks if the total payment for the winning workers is smaller than the current leftover budget (Step 26). If so, it employs the workers in the winning set to perform sensing tasks for the current round and subtracts their payments from the current budget (Steps 27-28). For each task that workers performed, the algorithm calculates the first ETD or second ETD by Eq. (14)-(18) (Steps 30-34). Besides, the algorithm will also update $n_{i,t}$, $\hat{r}_{i,t}$, $\hat{r}_{i,t}^+$ simultaneously based on the observed SR in the same current round (Step 36). At the end of each round, the algorithm will update $S_t^{HT}$ (Step 37). The algorithm alternates between the auction (i.e., exploitation) and the SR update (i.e., exploration) in each of the rounds until the remaining budget cannot afford the payments. So far, the algorithm returns the worker recruitment result and the payment profile.

### 4.3 Truthfulness, Individual Rationality and Efficiency

To prove the truthfulness, we reveal that workers can obtain the maximum utility if they bid truthfully.

**Theorem 1**. *The SCMABA mechanism is truthful.*

*Proof.* For $\forall$ worker $s_i \in \mathcal{S}$ submits an untruthful bid $b_{i,t}$ to complete tasks in $\mathcal{D}_{i,t}$, i.e., $b_{i,t} \neq c_{i,t}$. In the exploration phase, the task allocation is unrelated of the bid and the payment is fixed ($|\mathcal{D}_{i,t}|c_{max}$) when $s_i$ is recruited. Thus, the property is satisfied because the worker cannot increase its utility by manipulating the real cost. In the exploitation phase, in each round $t$, we denote the recruitment result for bidding $b_{i,t}$ and $c_{i,t}$ as $\varphi_{i,t}'$ and $\varphi_{i,t}$, respectively. There are two cases as below:

**Case 1 ($b_{i,t} > c_{i,t}$):** At this point, the worker falsely raises the bid, the worker is more difficult to be chosen by the platform, so we have $\varphi_{i,t}' \leq \varphi_{i,t}$.

If $\varphi_{i,t}' = \varphi_{i,t}$, as seen by $p_{i,t} = min\{\frac{\Sigma_{j|d_j \in \mathcal{D}_{i,t}} \omega_j \hat{r}_{i,t}^+}{\Sigma_{j|d_j \in \mathcal{D}_{a_{K+1,t}}} \omega_j \hat{r}_{a_{K+1,t}}^+} b_{a_{K+1,t}}, |\mathcal{D}_{i,t}|c_{max}\}$, the worker's payment ($p_{i,t}$) does not depend on his own bid ($b_{i,t}$), i.e. $p_{i,t}(b_{i,t}) = p_{i,t}(c_{i,t})$, and no higher gain is obtained.

If $\varphi_{i,t}' < \varphi_{i,t}$, the worker's false submission of the offer at this time resulted in the worker not being recruited by the platform, meaning that $\varphi_{i,t}' = 0$ and $\varphi_{i,t} = 1$. Obviously at this point, bidding truthfully is the better strategy bidding.

**Case 2 ($b_{i,t} < c_{i,t}$):** At this point, the worker falsely reduces the bid, making the worker more likely to be recruited by the platform, i.e., $\varphi_{i,t}' \geq \varphi_{i,t}$.

If $\varphi_{i,t}' = \varphi_{i,t}$, Similarly, since the worker's payment does not depend on his bid, the property is satisfied.

If $\varphi_{i,t}' > \varphi_{i,t}$, At this point, the worker falsely reports the bid, allowing the worker to be recruited by the platform, i.e., $\varphi_{i,t}' = 1$ and $\varphi_{i,t} = 0$. Since the platform selects the $K$ workers with the largest revenue-cost-ratio $\rho$ in each round, if worker $i$ is recruited, he needs to satisfy $\rho_{i,t} \geq \rho_{a_{K+1,t}}$, where $\rho_{i,t} = \left( \Sigma_{j|d_j \in \mathcal{D}_{i,t}} \omega_j \hat{r}_{i,t}^+ \right) / b_{i,t}$. This implies that $b_{i,t} \leq \frac{\Sigma_{j|d_j \in \mathcal{D}_{i,t}} \omega_j \hat{r}_{i,t}^+}{\rho_{a_{K+1,t}}}$ and $c_{i,t} \geq \frac{\Sigma_{j|d_j \in \mathcal{D}_{i,t}} \omega_j \hat{r}_{i,t}^+}{\rho_{a_{K+1,t}}}$. So we can derive $\frac{\Sigma_{j|d_j \in \mathcal{D}_{i,t}} \omega_j \hat{r}_{i,t}^+}{\rho_{a_{K+1,t}}} \leq c_{i,t} \leq |\mathcal{D}_{i,t}|c_{max}$, and thus the utility of worker $s_i$ is $u_{i,t}(b_{i,t}) = p_{i,t}(b_{i,t}) - c_{i,t} = min\left\{ \frac{\Sigma_{j|d_j \in \mathcal{D}_{i,t}} \omega_j \hat{r}_{i,t}^+}{\rho_{a_{K+1,t}}}, |\mathcal{D}_{i,t}|c_{max} \right\} - c_{i,t} \leq 0$. At this point, the worker's benefit is less than 0, and it is obvious that the truthful bid is the better strategy.

**Theorem 2**. *In each round $t$, the SCMABA mechanism is individually rational.*

*Proof.* In the $t$-th round, if a worker $s_i$ is not recruited, its utility is zero. Otherwise, if $s_i$ is recruited, its utility is $u_{i,t} = p_{i,t} - c_{i,t}$. In the exploration phase, the payment of $s_i$ is always $|\mathcal{D}_{i,t}|c_{max} \geq c_{i,t}$, so worker $s_i$'s utility is nonnegative. In the exploitation phase, if worker $s_i$ is recruited, we can derive that $b_{i,t} \leq \frac{\Sigma_{j|d_j \in \mathcal{D}_{i,t}} \omega_j \hat{r}_{i,t}^+}{\rho_{a_{K+1,t}}}$. Based on Theorem 1, each worker will bid truthfully, i.e., $b_{i,t} = c_{i,t}$. Hence, $c_{i,t} \leq \frac{\Sigma_{j|d_j \in \mathcal{D}_{i,t}} \omega_j \hat{r}_{i,t}^+}{\rho_{a_{K+1,t}}}$. Then, the utility of worker $s_i$ is $min\left\{ \frac{\Sigma_{j|d_j \in \mathcal{D}_{i,t}} \omega_j \hat{r}_{i,t}^+}{\rho_{a_{K+1,t}}}, |\mathcal{D}_{i,t}|c_{max} \right\} - c_{i,t} \geq 0$.

**Theorem 3**. *In each round $t$, the SCMABA mechanism is computationally efficient.*

*Proof.* The SCMABA mechanism consists of the exploration and exploitation phases. In each iteration of the while loop in the exploration phase, the computational complexity is at most $O(N)$ dominated by the quality update. In the exploitation phase, there are two main operations of sorting and payment, whose total computational complexity is $O(N \log N)$. The iteration of the recording bandit policy is an order of $O(K)$. In SCMABA, the number of while loops equals the total rounds $T$ bounded in $\frac{B}{Kc_{min}}$. Hence, SCMABA is computationally efficient.

## 5. Performance analysis

In this section, we conduct the performance evaluation of the SCMABA mechanism by using a real-world dataset.

### 5.1 Simulation setup

We conducted many simulation experiments with Chicago Taxi Trips [68] as the background. We consider drivers as sensing workers, and the platform will employ drivers to survey the traffic flow of a certain road segment at a certain moment. In contrast, a road segment at the same moment uniquely corresponds to a traffic counting task. Before performing tasks, drivers will select their preferred tasks and provide their quotes to the platform. After completing the task, they will report the task data to the platform.

Within the simulation experiment, we selected $\mathcal{M}$ road segments as sensing tasks and $\mathcal{N}$ drivers as sensing workers. With the number of tasks $\mathcal{M} = 40$, $\mathcal{N}$ is chosen from [40,140]. We define the number of workers selected in each round as $K$, which is chosen from {5,10,15,20}. In addition, the weight of tasks is uniform, and the actual cost of each task the drivers perform is related to the current ride, and the cost will not exceed the range $[c_{min}, c_{max}]$, while $c_{min} = 0.1$ and $c_{max} = 1$. Additionally, we assumed that the number of tasks a driver chosen to perform in each recruitment round is $|\mathcal{D}_{i,t}|$, taking values in [5,20] as a limitation given by the platform for practical reasons. Because a sensing worker has limited energy over a period of time and performing too many tasks over a period (e.g., in one day) may result in a decline in their sensing rate. Meanwhile, if he performs too few tasks, the platform has too little data available and the predicted completion rate will be inaccurate. Since there is no record of the workers' SRs, we use a split-bucket approach to generate the expected SRs of workers, with a stochastic distribution in each bucket. Specifically, 25% of workers with SRs in the interval [0.8, 1.0], 25% of workers with SRs in the interval [0.6, 0.8], 25% of workers with SRs in the interval [0.4, 0.6], 25% of workers with SRs in the interval [0, 0.4]. The simulation parameters are listed in Table 2.

### 5.2 Algorithms for comparison

Since the SCMABA mechanism takes into account the unknown SRs of workers and the incentive problems in MCS, no other existing algorithm can be directly applied to our problem, so we can only find algorithms that are close to our problem. Specifically, we selected the Random Algorithm and $\varepsilon$-Greedy Algorithm as the comparison algorithms.

In the Random Algorithm, in each round the platform randomly selects $K$ workers and pays them the maximum possible cost, i.e., $|\mathcal{D}_{i,t}|c_{max}$. In the $\varepsilon$-Greedy Algorithm, the SRs of workers is unknown and no longer follows the rule of exploration before exploitation. At each round, the platform will decide whether the round is for exploration or exploitation by $\varepsilon$, where $\varepsilon$ is chosen between [0,1]. Concretely, the platform has $\varepsilon$ probability of deciding this round as exploration, taking turns to select unknown workers to complete the task, evaluating their SRs by GTD and paying the maximum possible cost, and finally recording the SR information obtained in this phase. And the platform has $(1 - \varepsilon)$ probability of deciding this round as exploitation, where we considered that the $\varepsilon$-Greedy would select the top $K$ workers with the highest $\left(\sum_{j|d_j \in \mathcal{D}_{i,t}} \omega_j \hat{r}_{i,t}\right)/b_{i,t}$ among the workers that have been explored for recruitment and give a key payment to each worker who completes the task. In fact, when $\varepsilon$ is greater than 0.5, the SRs of workers has a tendency to explore. And when $\varepsilon$ is less than 0.5, it has a tendency to exploit. In order to fully characterize the algorithm in this simulation and show its performance in terms of exploration and exploitation, we choose two values of $\varepsilon$ as 0.3 and 0.7 for the simulation, denoted as 0.3-Greedy and 0.7-Greedy, respectively.

### 5.3 Evaluation results

We first analyze the total revenue and regret of different algorithms in the same situation and compare our SCMABA with other comparative algorithms to prove the outstanding performance of SCMABA, where regret is the difference between the revenue obtained by the proposed algorithms and the optimal solution. The results show that our SCMABA algorithm exhibits significantly better performance than the comparison algorithms in terms of total revenue and regret in all cases. Specifically, the total gain obtained by our algorithm is at least 9.34% higher than the other three algorithms, and the regret produced by our algorithm is at least 18.59% lower than the other algorithms. Then, we explore the algorithms' persistence, truthfulness, individual rationality, and worker identification. For one thing, it is revealed that the SCMABA can run longer than the other three algorithms within the same budget, and it satisfies the requirements of authenticity and individual rationality. For another, we also found that our SCMABA has excellent performance for the identification of workers' SRs.

First, we explore the effect in total budget $B$ on total revenue and regret, all else being equal. We conducted multiple experiments for $B$ from 6000 to 12000 with each increment of 1000, and the experimental results are shown in Figs. 5-6. It is observed in Fig. 5 that the SCMABA has the highest revenue, followed by 0.3-Greedy and 0.7-Greedy, and the lowest is the Random Algorithm. Statistically, the SCMABA achieves about 28.11% higher revenue than 0.3-Greedy, and even three times the revenue of the Random Algorithm. In the $\varepsilon$-Greedy algorithm, the platform has a higher probability of choosing to explore in the algorithm with a larger $\varepsilon$ value, and the expense of exploration will be higher than the exploitation, resulting in a lower total execution round. So, the gain of 0.3-Greedy is greater than 0.7-Greedy and regret is less than 0.7-Greedy. The Random Algorithm recruits workers randomly and therefore has the lowest payoff and highest regret. With the increase of total budget $B$, the platform can recruit more workers, and thus complete more tasks, so the revenue of all algorithms is increased. In Fig. 6 we can observe that the regret of a fifth of the regret of the Random Algorithm. Similarly, as the total budget $B$ increases, the platform can recruit more workers to complete more tasks, so the regret of all algorithms increases, and the SCMABA algorithm has the slowest growth rate of regret.

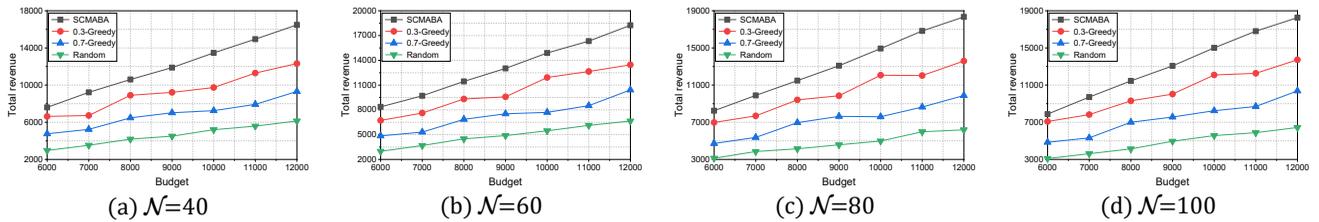

**Fig. 5: Total revenue vs. budget $B$**

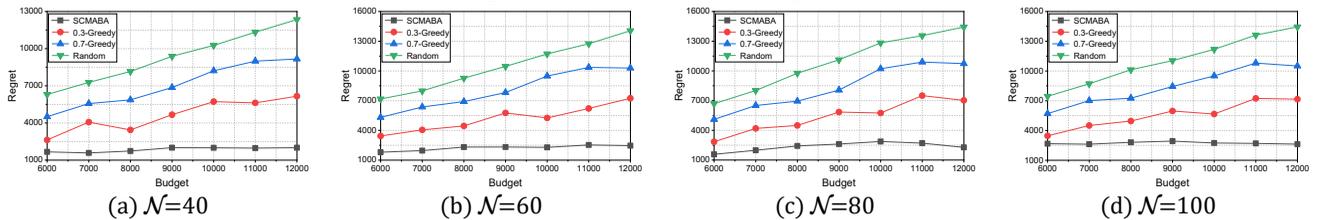

**Fig. 6: Regret vs. budget $B$**

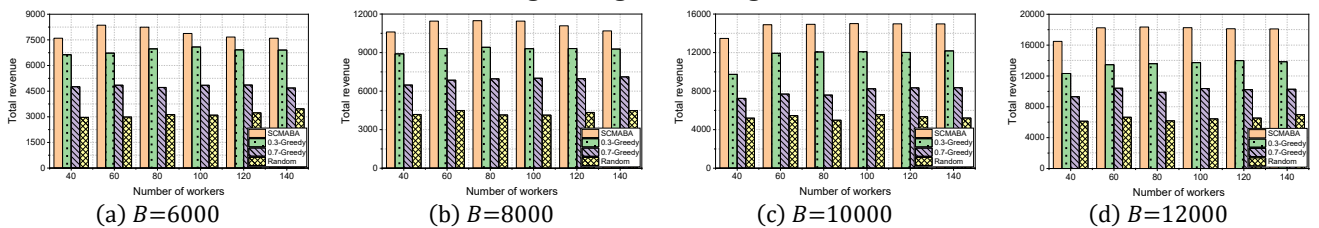

**Fig. 7: Total revenue vs. number of workers $\mathcal{N}$**

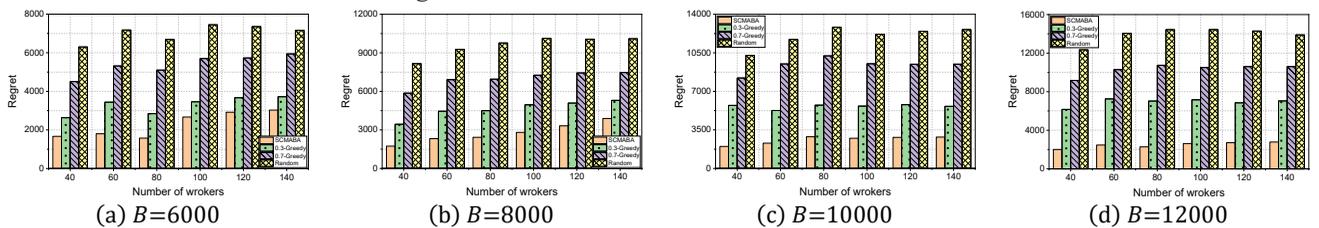

**Fig. 8: Regret vs. number of workers $\mathcal{N}$**

TABLE 2: Simulation settings

| Parameter name | Values |
| --- | --- |
| Budget $B$ | $6, 7, 8, 9, 10, 11, 12 \times 10^3$ |
| number of tasks $\mathcal{M}$ | 40 |
| preferred tasks $\mathcal{D}_{i,t}$ | [5,15] |
| number of worker $\mathcal{N}$ | 40, 60, 90, 100, 120, 140 |
| number of selected workers $K$ | 5, 10, 15, 20 |
| range of cost $[c_{min}, c_{max}]$ | [0.1, 1] |

Then, we explored the effect of the total number of sensing workers $\mathcal{N}$ on total revenue and regret. We conducted experiments for $\mathcal{N}$ from 40 to 140 with each increment of 20, and the results are shown in Figs. 7-8. The data shows that the SCMABA gets more revenue and lower regrets than other algorithms. It can be easily observed that the SCMABA outperforms the other comparison algorithms in terms of both revenue and regret. With a small budget, as the number of workers increases, the revenue received gradually decreases and the regret gradually increases. The reason for this phenomenon is that as the number of workers increases, too high a percentage of the budget is spent on the exploration process. With a large budget, as the number of workers grows, the revenue obtained by the SCMABA increases and the regret decreases. This is because as the number of workers grows, there are more highly trustworthy workers with lower cost in the system, resulting in better performance of the SCMABA.

Next, we explored the effect of the number of workers selected in each round $K$ on total revenue and regret, and the experimental results are shown in Figs. 9-10. Under different $K$ values, the SCMABA still performs better than other comparative algorithms. Meanwhile, with the $K$ value increasing, the total revenue of both the SCMABA and the $\varepsilon$-Greedy Algorithms decreased, and their total regret both increased. Statistically, comparing Fig. 9 (a) with Fig. 9 (d), it can be found that the revenue of the SCMABA algorithm decreases by 17.17%. Similarly, comparing Fig. 9 (a) and Fig. 9 (d), it is obvious that the regret of SCMABA algorithm increases by 71.18%. This is because, as $K$ increases, the platform needs to recruit more suboptimal workers in each round, and the suboptimal workers will bring lower revenue.

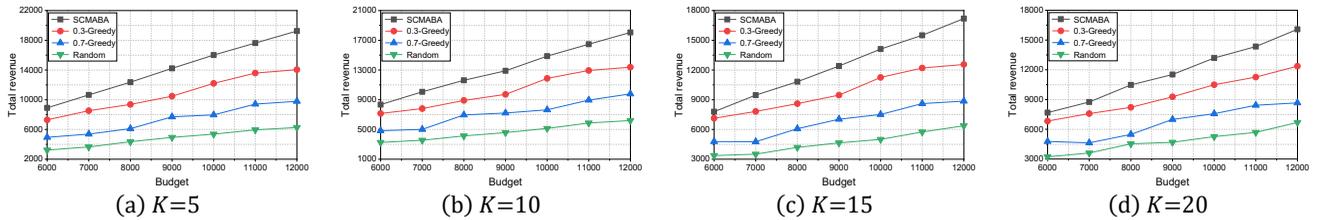

(a) $K$=5  (b) $K$=10  (c) $K$=15  (d) $K$=20

**Fig. 9: Total revenue vs. number of selected workers $K$**

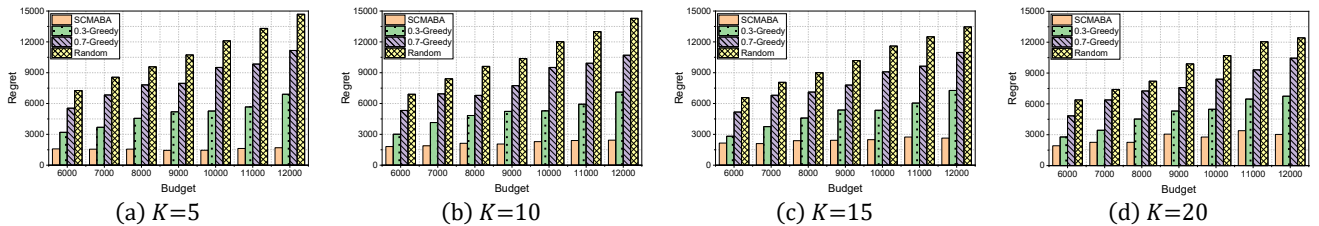

(a) $K$=5  (b) $K$=10  (c) $K$=15  (d) $K$=20

**Fig. 10: Regret vs. number of selected workers $K$**

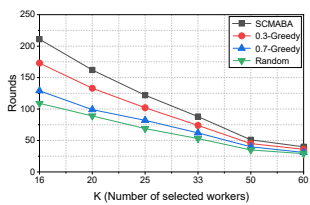
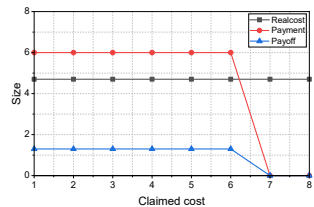
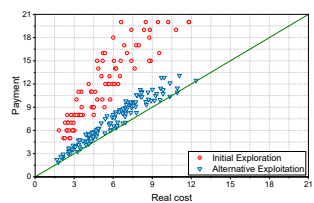
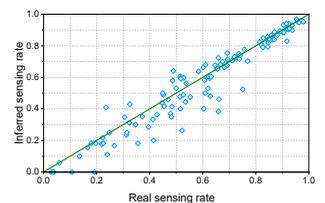

**Fig. 11:Total rounds vs. $K$**  **Fig. 12: Truthfulness**  **Fig. 13: Individual rationality**  **Fig. 14: Worker identification**

Last but not least, to explore the SCMABA algorithm's properties, we considered four measures: persistence, truthfulness, individual rationality, and worker identification.

*Persistence*. We hold $\mathcal{N}$ constant at 100 and explore the number of rounds that the SCMABA lasts for different values of $K$ and compare it with three other comparative algorithms. As can be seen in Fig. 11, SCMABA lasts longer in all cases. And the advantage of SCMABA is more obvious when the value of $K$ is less. Specifically, when $K$=16, SCMABA can run 21.96% longer than 0.3-Greedy Algorithm, and the running time is almost twice as long as Radom Algorithm.

*Truthfulness*. We randomly selected a winning bidder, changed its claimed cost, and recalculated the payment and utility. It is shown in Fig. 12 that if the claimed cost does not exceed the payment, the utility remains constant. However, when the claimed cost is greater than the threshold, the utility becomes zero. Therefore, the SCMABA mechanism satisfies truthfulness.

*Individual rationality*. In Fig. 13, we calculate the workers' payments in the exploration and exploitation phases respectively under the conditions of $\mathcal{N}=120$ and $K=40$. The figure shows that each payment is no less than the corresponding actual cost, which indicates that SCMABA satisfies individual rationality. At the same time, we can find that the payment in the exploration phase is significantly larger than the corresponding actual cost, while the payment in the exploitation phase is relatively close to the actual cost. This is because during the exploration phase workers were paid $c_{max}$, most of which was much higher than the cost. While the payment in the exploitation phase is calculated and is close to the actual cost.

*Worker identification*. In Fig. 14, we explore the identification performance of the workers' SRs of the SCMABA. Overall, SCMABA is accurate in the estimation of workers' SRs. Also, we can find that the SCMABA algorithm is more accurate for the estimation of workers with higher actual SRs. This is because workers with higher actual SRs are recruited more often and are evaluated more adequately.

# 6. Conclusion and future work

In this paper, we focus on the worker recruitment problem in MCS while considering the unknown SRs and strategic workers simultaneously. First of all, we model the worker recruitment problem as a novel $K$-armed combinatorial multi armed bandit problem and regard workers' SRs as the gain. Then, we adopt reverse auction to incentivize workers' participation as well as discourage their strategic behaviors. Moreover, a SR acquisition mechanism named SSRL is proposed to quickly and accurately obtain the workers' SRs. Combining the SR acquisition mechanism with multi-armed bandit reverse auction, we designed SCMABA, which solves the recruitment problem of multiple unknown and strategic workers in MCS. Finally, we conduct extensive simulations on a real dataset to demonstrate the effectiveness of our mechanisms.

The SCMABA algorithm utilizes algorithmic game theory and auction theory to ensure truthfulness, individual rationality, and computational efficiency, providing theoretical guarantees for feasibility. It also uses the UCB algorithm to balance exploration and exploitation to maximize the benefits of the platform, which solves the Information Elicitation Without Verification (IEWV) problems and can be easily generalized. But the Semi-supervised SR Learning algorithm is specifically designed to evaluate the data quality in MCS and cannot be directly applied to the quality evaluation in other nodes. If the SCMABA scheme is to be applied to other networks, different quality evaluation algorithms adapted to the actual scenario need to be designed for the different requirements of each network.

In the future, we will devote ourselves to further broadening the boundaries of MAB modeled MCS based on the auction. Specifically, we will further consider the location of tasks and the trajectory of workers in the auction process, to balance the assignment of workers to reduce the cost of MCS and enhance social welfare. Additionally, we will also aim to generalize the SCMABA solution to other Unknown Node Selection (UNS) problems in various types of networks.

## CRediT authorship contribution statement

**Jianheng Tang**: Conceptualization, Methodology, Software, Investigation, Formal analysis, Writing-original draft. **Kejia Fan**: Methodology, Formal analysis, Writing-original draft, Review & editing. **Wenxuan Xie**: Software, Review & editing. **Luomin Zeng**, **Feijiang Han**: Review & editing. **Guosheng Huang** and **Tian Wang**: Supervision, Review & editing. **Anfeng Liu**: Conceptualization, Funding acquisition, Resources, Supervision, Writing-original draft, Writing-review & editing. **Shaobo Zhang**: Supervision, Review & editing.

## Declaration of competing interest

The authors declare that they have no known competing financial interests or personal relationships that could have appeared to influence the work reported in this paper.

## Data availability

Data will be made available on request.

## Acknowledgement

This work was supported in part by the National Natural Science Foundation of China (62072475).

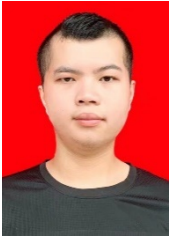

**Jianheng Tang** is currently a student at the School of Computer Science and Engineering, Central South University, China. His research interests include mobile crowdsensing, reinforcement learning, and incentive mechanism.

E-mail: tangent-heng@csu.edu.cn.

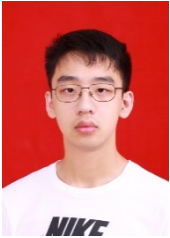

**Kejia Fan** is currently a student at the School of Computer Science and Engineering, Central South University, China. His research interests include crowdsourcing, auction theory, and incentive mechanism.

E-mail: kejiafan@csu.edu.cn.

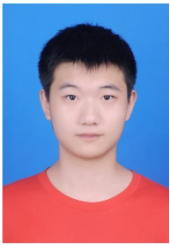

**Wenxuan Xie** is currently a student at the School of Computer Science and Engineering, Central South University, China. His research interests include auction theory, crowdsensing, and reinforcement learning.

E-mail: 8208200422@csu.edu.cn.

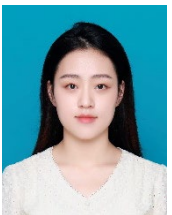

**Luomin Zeng** is currently a student at the School of civil engineering, Central South University, China. Her research interests include mobile crowd sensing and incentive mechanism.

E-mail: luominzeng@csu.edu.cn.

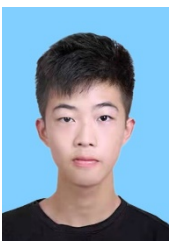

**Feijiang Han** is currently a student at the School of Computer Science and Engineering, Central South University, China. His research interests include auction theory, incentive mechanism, and reinforcement learning.

E-mail: feijianghan@gmail.com.

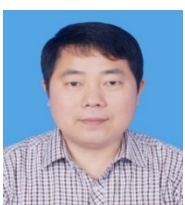

**Guosheng Huang** received his M.S. and Ph.D. degrees in computer science from Central South University, China, 2001 and 2010 respectively. Currently, he is an associate professor of School of Information Science and Engineering, Hunan First Normal University, China. His major research interest is mobile computing.

Email: huanggs@hnfnu.edu.cn.

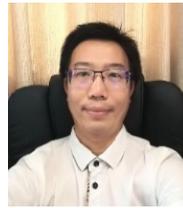

**Tian Wang** received his BSc and MSc degrees in Computer Science from Central South University in 2004 and 2007. He received his Ph.D. degree at the City University of Hong Kong in 2011. Currently, he is a professor at the Artificial Intelligence and Future Networks, Beijing Normal University & UIC, China. His research interests include the internet of things, edge computing, and mobile computing.

Email: tianwang@bnu.edu.cn.

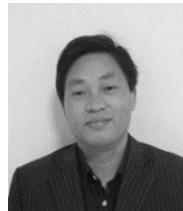

**Anfeng Liu** received the M.Sc. and Ph.D. degrees from Central South University, China, in 2002 and 2005, respectively, both in computer science. He is currently a professor of the School of Information Science and Engineering, Central South University, China. His major research interest is wireless sensor networks, Internet of Things, information security, edge computing and crowdsourcing. Dr. Liu has published 4 books and over 200 international journal and conference papers, among which there are more than 30 ESI highly-cited papers. Some of his works were published in IEEE Transactions on Information Forensics & Security, IEEE Transactions on Mobile Computing, IEEE Transactions on Parallel and Distributed Systems, IEEE Transactions on Vehicular Technology, IEEE Transactions on Computer-Aided Design of Integrated Circuits and Systems. His research has been supported by the National Basic Research Program of China (973 Program) and the National Natural Science Foundation of China for five times. He was a recipient of the First Prize of Scientific Research Achievement of Colleges from the Ministry of Education of China in 2016, and the Second Prize of Science and Technology Award from China Nonferrous Metal Industry Association in 2005. He has served as the Editor of the IEEE Networking Letters.

E-mail: afengliu@mail.csu.edu.cn.

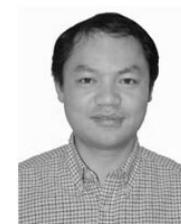

**Shaobo Zhang** received the B.Sc. and M.Sc. degree in computer science both from Hunan University of Science and Technology, Xiangtan, China, in 2003 and 2009 respectively, and the Ph.D. degree in computer science from Central South University, Changsha, China, in 2017. He is currently an associate professor at School of Computer Science and Engineering of the Hunan University of Science and Technology, China. His research interest is cloud computing.

E-mail: shaobozhang@hnust.edu.cn.